\documentclass{amsart}

\usepackage{epsfig}

\allowdisplaybreaks[1]
\numberwithin{equation}{section}
\newcommand{\Li}[1]{{\rm Li}_{#1}}   
\newcommand{\Cl}[1]{{\rm Cl}_{#1}}

\newcommand\half{\tfrac{1}{2}}

\newcommand{\abs}[1]{\lvert#1\rvert}

\renewcommand{\Im}{{\rm Im}}

\begin{document}

\def\check{\marginpar{\fbox{check!}}}

\def\dd{{\rm d}}

\newcommand{\Int}{\int\limits}

\title[Infinite series]{Some infinite series related to Feynman diagrams}

\author{Odd Magne Ogreid}
\address{Department of Physics \\  University of Bergen, All\'egt.~55 \\
N-5007 Bergen\\ Norway}
\email{Odd.Ogreid@fi.uib.no}
\urladdr{http://www.fi.uib.no/\~{}ogreid/}
\author{Per Osland}
\address{Department of Physics \\  University of Bergen, All\'egt.~55 \\
N-5007 Bergen\\ Norway}
\email{Per.Osland@fi.uib.no}
\urladdr{http://www.fi.uib.no/\~{}osland/particle.html}

\thanks{This research has been supported by the Research Council of Norway}

\begin{abstract}
Results are presented for some infinite series appearing in Feynman diagram
calculations, many of which are similar to the Euler series.
These include both one-, two- and three-dimensional series.
The sums of these series can be evaluated with the help of various
integral representations for hypergeometric functions, and expressed in 
terms of $\zeta(2)$, $\zeta(3)$, the Catalan constant $G$ and 
${\rm Cl}_2(\pi/3)$ where ${\rm Cl}_2(\theta)$ is Clausen's function.
\end{abstract}
\maketitle

\section{Introduction}
\label{s:intro}

High-precision calculations in Quantum Field Theory require
the evaluation of Feynman amplitudes, which represent quantum
corrections in a perturbative expansion.
This expansion is governed by a small parameter, the fine-structure
constant, and related to the number of loops in the associated
Feynman diagrams.
These are graphical devices to aid in the book-keeping.

The evaluation of such Feynman amplitudes can technically be rather 
difficult, and one must often resort to approximate methods, valid in 
certain kinematical limits. For the case of electron-positron scattering,
the full second-order corrections are not yet known.
However, in the limit of high energies and small scattering angles,
the problem simplifies: the relevant part of the so-called
virtual amplitude can be expanded in powers of large logarithms.
The coefficients and arguments of these logarithms can be extracted
by the use of suitable Mellin transforms.
We shall here review some of the techniques involved,
and present some of the mathematical results of such calculations.

\section{Physics Problem}
\label{s:physics}
The mathematical results we will discuss, derive from the ``two-loop''
Feynman diagram depicted in Fig.~1.
The solid lines represent an electron and a positron scattering
via the exchange of photons (wiggly lines).
\begin{figure}[htb]
\refstepcounter{figure}
\label{Fig:double-box}
\addtocounter{figure}{-1}
\begin{center}
\setlength{\unitlength}{1cm}
\begin{picture}(3.5,3.5)
\put(-3,0.0)
{\mbox{\epsfysize=3.5cm\epsffile{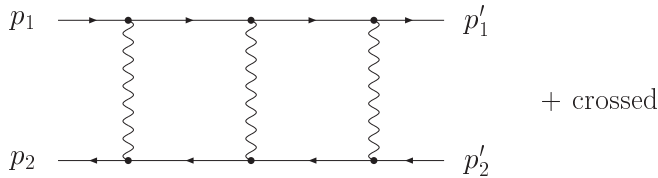}}}
\end{picture}
\caption{Double-box diagram}
\end{center}
\end{figure}
The momenta $p_1$, $p_2$, $p_1'$ and $p_2'$ only enter through
the combinations
\begin{equation}
s=(p_1+p_2)^2, \quad t=(p_1-p_1')^2=(p_2-p_2')^2, \quad
m^2=p_i^2=p_i'{}^2 
\end{equation}
where $s$ and $-t$ represent squares of the over-all energy
and momentum transfer, respectively.
Furthermore, $m$ is the electron mass.
The corresponding Feynman amplitude is proportional to
the integral
\begin{equation}
\label{Eq:feynman}
M=\int\frac{\dd^7\alpha\; \delta(1-\sum\alpha_k) \Lambda^{-3}(\alpha)}
   {\left[D_s(\alpha)s+D_t(\alpha)t+D_m(\alpha)m^2
  +D_\lambda(\alpha)\lambda^2\right]^3}
\end{equation}
where a fictitious photon mass, $\lambda$, has been introduced
in order to make the integrand well-defined for all $\alpha_j$.
The $D_i(\alpha)$ and $\Lambda(\alpha)$ are homogeneous functions 
of seven $\alpha_j$,
one associated with each internal line in the diagram.

Asymptotically, for $s\gg |t|\gg m^2\gg \lambda^2$ 
(high energies, small angles), the integral $M$ has the structure
\begin{equation}
\label{Eq:asymptotic}
M\simeq A_4\log^4(\cdots)+A_3\log^3(\cdots)+\cdots
\end{equation}
where the arguments of the logarithms are given by the kinematical
variables $s$, $t$, $m^2$ and $\lambda^2$.
\section{Mathematical Approach}
\label{s:approach}
\subsection{Mellin Transform}
In order to extract, from the integral (\ref{Eq:feynman}), 
the asymptotic behaviour given by Eq.~(\ref{Eq:asymptotic}), 
we perform a Mellin transform.
If the function $f(x)$ behaves ``almost'' like $x^{-l}$ at large $x$,
then we define the Mellin transform as
\begin{equation}
J(z)=\Int_0^\infty x^{l-z}\, f(x) \dd x
\end{equation}
The fact that $x^{l-z}\,f(x)$ is only marginally integrable,
is represented by poles in $J(z)$, as $z\to0$.
The inverse transform is given by:
\begin{equation}
f(x)=\frac{1}{x^{l+1}}\,
\frac{1}{2\pi i}\Int_{c-i\infty}^{c+i\infty} x^z\, J(z) \dd z, \quad
c>0
\end{equation}
where the contour of integration is as depicted in Fig.~2.
\begin{figure}[htb]
\refstepcounter{figure}
\label{Fig:Mellin}
\addtocounter{figure}{-1}
\begin{center}
\setlength{\unitlength}{1cm}
\begin{picture}(6.0,6.0)
\put(-1.0,0.0)
{\mbox{\epsfysize=6.0cm\epsffile{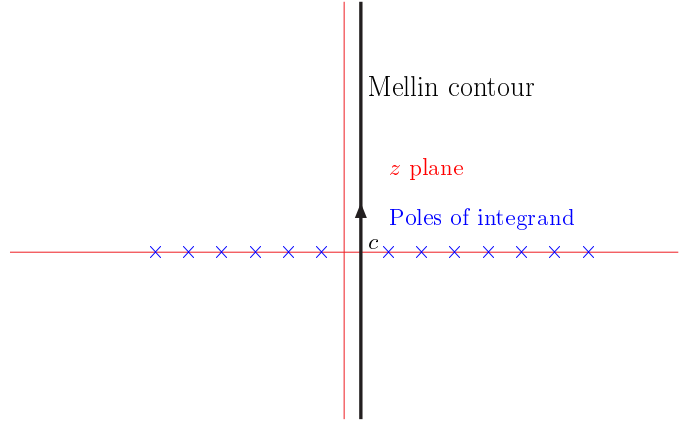}}}
\end{picture}
\caption{Contour used for the Inverse Mellin transform}
\end{center}
\end{figure}

It is instructive to consider the simple example
\begin{equation}
f(x)=\theta(x-1)\,\frac{1}{x^{l+1}}\,\log^k x,
\end{equation}
for which the Mellin transform gives
\begin{equation}
J(\zeta)=\frac{k!}{\zeta^{k+1}}
\end{equation}
Thus, an expansion of the Mellin transform in inverse powers of $\zeta$ 
is equivalent to an expansion in powers of logarithms.
\subsection{Factorization Formula}
In order to facilitate the integration over the $\alpha_j$ in 
(\ref{Eq:feynman}), we invoke the ``factorization formula''
\cite{BW,OW-86}.
For three terms, it can be written as:
\begin{eqnarray}
\label{Eq:factorization}
&&(A_1+A_2+A_3)^{-p} \nonumber \\
&&=\int_{c_1}\frac{\dd z_1}{2\pi i}
 \int_{c_2}\frac{\dd z_2}{2\pi i}\,
\frac{\Gamma(z_1)\Gamma(z_2)\Gamma(p-z_1-z_2)}{\Gamma(p)} 
A_1^{-z_1}\, A_2^{-z_2}\, A_3^{-p+z_1+z_2} \\[4pt]
&&\quad\Im\ A_i>0, \quad c_j>0, \quad p-c_1-c_2>0 \nonumber 
\end{eqnarray}
where the integration contours labelled $c_j$ run from 
$c_j-i\infty$ to $c_j+i\infty$.
To prove the factorization formula, one will need two preliminary
results. The first one is
\begin{eqnarray}
u^{-p}=\frac{e^{-i\pi p/2}}{\Gamma(p)}\int_0^\infty t^{p-1}e^{itu}{\rm d}t,
\hspace{1.5cm} {\rm Re\ }p>0,\quad {\rm Im\ }u>0.\label{firstresult}
\end{eqnarray}
{\em{Proof:}} First, change variables by putting $x=-itu$.
Using the integral representation of the Gamma function, we find,
\begin{eqnarray}
\frac{e^{-i\pi p/2}}{\Gamma(p)}\int_0^{-iu\infty}(-iu)^{-p}
e^{-x}x^{p-1}{\rm d}x=\frac{e^{-i\pi p/2}}{\Gamma(p)}e^{i\pi p/2}u^{-p}
\Gamma(p)=u^{-p}\nonumber
\end{eqnarray}
which is valid if $u$ is purely imaginary with $u/i>0$ and Re $p>0$. This
result can be extended to any $u$ with Im $u>0$ by analytic continuation.
The second result needed is
\begin{eqnarray}
e^{iT}=\frac{1}{2\pi i}
\int_{c}\Gamma(z)e^{i\pi z/2}T^{-z}{\rm d}z,\hspace{1.5cm}
c>0,\quad {\rm Im\ }T>0.
\end{eqnarray}
{\em{Proof:}} Consider the Mellin transform of $e^{iTx}$
[cf. Eq.~(\ref{firstresult})],
\begin{eqnarray}
I^\prime&=&\int_0^\infty x^{z-1}e^{iTx}{\rm d}x=\Gamma(z)e^{i\pi z/2}T^{-z},
\nonumber
\end{eqnarray}
which is valid for Re $z>0$ and Im $T>0$. By taking the inverse Mellin
transform of this expression, one should get $e^{iTx}$,
\begin{eqnarray}
e^{iTx}&=&\frac{1}{2\pi i}
\int_{c^{\prime}}x^{-z}\Gamma(z)e^{i\pi z/2}T^{-z}{\rm d}z.\nonumber
\end{eqnarray}
By putting $x=1$ in this expression, the desired result is obtained.
One is now capable of proving the factorization formula.\\
{\em{Proof of the factorization formula:}}
\begin{eqnarray}
(A_1+A_2+A_3)^{-p}&=&\frac{e^{-i\pi p/2}}{\Gamma(p)}\int_0^\infty{\rm d}t
t^{p-1}e^{it(A_1+A_2+A_3)}\nonumber\\
&=&\frac{e^{-i\pi p/2}}{\Gamma(p)}\int_0^\infty{\rm d}t t^{p-1}
e^{itA_1}e^{itA_2}e^{itA_3}\nonumber\\
&=&\frac{e^{-i\pi p/2}}{\Gamma(p)}\int_0^\infty{\rm d}t t^{p-1}
\int_{c_1}\frac{{\rm d}z_1}{2\pi i}\int_{c_2}\frac{{\rm d}z_2}{2\pi i}
\Gamma(z_1)\Gamma(z_2)\nonumber\\&&\times
e^{i\pi z_1/2}e^{i\pi z_2/2}
(tA_1)^{-z_1}(tA_2)^{-z_2}
e^{itA_3}\nonumber\\
&=&\frac{e^{-i\pi p/2}}{\Gamma(p)}
\int_{c_1}\frac{{\rm d}z_1}{2\pi i}\int_{c_2}\frac{{\rm d}z_2}{2\pi i}
\Gamma(z_1)\Gamma(z_2)\nonumber\\&&\times
e^{i\pi z_1/2}e^{i\pi z_2/2}
A_1^{-z_1}A_2^{-z_2}
\int_0^\infty{\rm d}t
t^{p-1-z_1-z_2}e^{itA_3}\nonumber\\
&=&\frac{e^{-i\pi p/2}}{\Gamma(p)}
\int_{c_1}\frac{{\rm d}z_1}{2\pi i}\int_{c_2}\frac{{\rm d}z_2}{2\pi i}
\Gamma(z_1)\Gamma(z_2)
e^{i\pi z_1/2}e^{i\pi z_2/2}\nonumber\\&&\times
A_1^{-z_1}A_2^{-z_2}
\Gamma(p-z_1-z_2)
e^{i\pi(p-z_1-z_2)/2}A_3^{-p+z_1+z_2},\nonumber
\end{eqnarray}
from which the factorization formula follows. The formula is 
valid for Im $A_i>0$, $c_i>0$ and
$p-c_1-c_2>0$.
It is trivially generalized to $(A_1+A_2+\cdots A_n)^{-p}$.
\subsection{Emergence of Sums}
Performing a Mellin transform in $s$, with
$|t|$=1, $m^2=s^{\eta_m}$, $\lambda^2=s^{\eta_\lambda}$, 
and $\eta_\lambda < \eta_m < 0$,
one is led to integrals over $z_i$ of the kind given
by Eq.~(\ref{Eq:factorization}).
These integrals are most conveniently performed by the use of
Cauchy's theorem, closing the contour in the left or right half-plane.
Expanding these summands in powers of the Mellin transform variable,
one is led to sums involving, as residues, the $\Gamma$ function
and its derivatives,
\begin{equation}
\psi(z)=({\rm d}/{\rm d}z)\, \log\Gamma(z)=\Gamma'(z)/\Gamma(z) 
\nonumber
\end{equation}
and
\begin{equation}
\psi'(z)=({\rm d}/{\rm d}z)\,\psi(z).
\nonumber
\end{equation}
The combination $\gamma+\psi(n)$ frequently appears in the summand, 
and several series of this type are evaluated using the following integral 
representation,
\begin{equation}
\gamma+\psi(z)=\int_0^1{\rm d}t\, \frac{1-t^{z-1}}{1-t}. \nonumber
\end{equation}
Here, $\gamma=0.577\; 216\dotsc$ is Euler's constant. 
\subsection{Special Functions}
Our evaluation of the sums makes use of several special functions.
We shall list a few of them along with some of their properties here:\\
\em{Riemann's zeta function:}\em\\
The sums of several series can be expressed in terms of the Riemann $\zeta$ 
function:
\begin{equation*}
\zeta(z)=\sum_{k=1}^\infty\frac{1}{k^z}
\end{equation*}
with
\begin{align*}
&\sum_{k=1}^\infty\frac{1}{k^2}=\zeta(2)=\frac{\pi^2}{6}
=1.644\; 934 \dots,\\
&\sum_{k=1}^\infty\frac{1}{k^3}=\zeta(3)
=1.202\; 057 \dots.
\end{align*}\\
\em{Polylogarithms:}\em\\
Furthermore, we need Euler's polylogarithms:
\begin{equation}
\Li{n}(z)=\sum_{k=1}^\infty\frac{z^k}{k^n}, \quad
\abs{z}<1,\quad n=0,1,2,\dotsc, \nonumber
\end{equation}
These functions also have integral representations:
\begin{align}
\Li{n}(z)&=\frac{(-1)^{n-1}}{(n-2)!}\int_0^1{\rm d}t\, 
\frac{\log^{n-2}(t)\log(1-tz)}{t}, \qquad
n=2,3,4,\dotsc  \\
&=\int_0^z{\rm d}t\, \frac{\Li{n-1}(t)}{t},\quad n=1,2,3,\dotsc,
\end{align}
with
$\Li{n}(1)=\zeta(n),\ n=2,3,4,\dotsc$,
$\Li{n}(-1)=(1/2^{n-1}-1)\zeta(n),\ n=2,3,4,\dotsc$.\\
\em{Clausen's function:}\em\\
On the unit circle, the imaginary part of the dilogarithm is
Clausen's function:
\begin{equation}
\Cl{2}(\theta)=\Im\bigl[\Li2(e^{i\theta})\bigr]
=\sum_{k=1}^\infty\frac{\sin k\theta}{k^2}. \nonumber
\end{equation}
Clausen's function has its maximum value for $\theta=\pi/3$, 
$\Cl{2}(\pi/3)=1.014\; 942\dotsc$.
Furthermore, $\Cl{2}(\pi/2)=G=0.915\; 966\dotsc$, where $G$ is Catalan's 
constant.
\subsection{A Theorem For Summing Series}
Many of the results given here were obtained using ``Theorem 1'' 
of \cite{OgreidOsland-98}:
\begin{eqnarray}
\sum_{n=1}^\infty\frac{1}{n^2}[\gamma+\psi(1+kn)] 
&\!=\!&\left(\frac{k^2}{2}+\frac{3}{2k}\right)\zeta(3)
+\pi\sum_{j=1}^{k-1}j\, \Cl{2}\left(\frac{2\pi j}{k}\right) \\[4mm]
\qquad \sum_{n=1}^\infty\frac{(-1)^n}{n^2}[\gamma+\psi(1+kn)] 
&\!=\!&\left(\frac{k^2}{2}-\frac{9}{8k}\right)\zeta(3)
+\pi\sum_{j=1}^{k-1}j\, \Cl{2}\left(\frac{2\pi j}{k}+\frac{\pi}{k}\right)
\end{eqnarray}
for $k=1,2,3,\dotsc$. The sums over $j$ are understood to be zero when $k=1$.

To prove the theorem, we use properties of Nielsen's generalized polylogarithms
\cite{Kolbig}. For a detailed proof of the theorem, we refer to 
\cite{OgreidOsland-98}.

\section{One-dimensional series}
\label{s:one-d}
We present a collection of one-dimensional series and their sums. For some 
of the series we also present the methods used to sum them. 
For the rest of the series, the proofs can be found in 
\cite{OgreidOsland-98,OgreidOsland-99}.
\begin{align}
&\sum_{n=1}^\infty\frac{1}{n^2}[\gamma+\psi(1+n)]=2\zeta(3)\\  
&\sum_{n=1}^\infty\frac{1}{n^2}[\gamma+\psi(1+2n)]=\frac{11}{4}\zeta(3)\\ 
&\sum_{n=1}^\infty\frac{1}{n^2}[\gamma+\psi(1+3n)]=
5\zeta(3)-\frac{2\pi}{3}\Cl{2}\left(\frac{\pi}{3}\right)\\
&\sum_{n=1}^\infty\frac{1}{n^2}[\gamma+\psi(1+4n)]=
\frac{67}{8}\zeta(3)-2\pi G\\
&\sum_{n=1}^\infty\frac{1}{n^2}[\gamma+\psi(1+6n)]=
\frac{73}{4}\zeta(3)-\frac{16\pi}{3}\Cl{2}\left(\frac{\pi}{3}\right)
\end{align}
Some of these series were known already by Euler!
Alternating series give similar results:
\begin{align}
&\sum_{n=1}^\infty\frac{(-1)^n}{n^2}[\gamma+\psi(1+n)]=-\frac{5}{8}\zeta(3)\\
&\sum_{n=1}^\infty\frac{(-1)^n}{n^2}[\gamma+\psi(1+2n)]=
\frac{23}{16}\zeta(3)-\pi G\\ 
&\sum_{n=1}^\infty\frac{(-1)^n}{n^2}[\gamma+\psi(1+3n)]=
\frac{33}{8}\zeta(3)-2\pi\Cl{2}\left(\frac{\pi}{3}\right)
\end{align}
The results presented so far, all follow from the theorem or the immediate 
corollary thereof. 

Some series encountered had to be summed by other means:
\begin{eqnarray}
&&\sum_{n=1}^\infty\frac{1}{n(n+1)}[\gamma+\psi(1+n)]=\zeta(2)\label{s7}\\
&&\sum_{n=1}^\infty\frac{1}{n^2(n+1)}[\gamma+\psi(1+n)]=2\zeta(3)-\zeta(2)
\label{s13} \\
&&\sum_{n=1}^\infty\frac{1}{n(n+1)^2}[\gamma+\psi(1+n)]=-\zeta(3)+\zeta(2)
\label{s10}
\end{eqnarray}
Related sums, where the argument of the psi function is shifted
by a small integer, are easily obtained using the recursion relation
$\psi(1+n)=\psi(n)+1/n$.\\
\em Proof of (\ref{s7}):\em \ We apply the integral representation of the
psi function before summing over $n$:
\begin{eqnarray}
&&\sum_{n=1}^\infty\frac{1}{n(n+1)}[\gamma+\psi(1+n)]
=\sum_{n=1}^\infty\frac{1}{n(n+1)}\int_0^1{\rm d}t\frac{1-t^n}{1-t}
\nonumber\\
&&=\int_0^1{\rm d}t\frac{1}{1-t}\left[1-\frac{t}{2}\ {}_2F_1(1,1;3;t)\right]
\nonumber\\
&&=\int_0^1{\rm d}t\frac{1}{1-t}
\left\{1-\frac{1}{t}\left[t+(1-t)\log(1-t)\right]\right\} \nonumber\\
&&=-\int_0^1{\rm d}t\frac{\log(1-t)}{t}=\zeta(2)
\nonumber
\end{eqnarray}
Here, we used (7.3.2.150) of Prudnikov \cite{Prudnikov}
to rewrite ${}_2F_1$. The resulting 
integral is just the definition of $\mbox{Li}_2(1)$ which is well known.

Series containing the derivative of the psi function were also encountered:
\begin{eqnarray}
&&\sum_{n=1}^\infty\frac{1}{n}\psi^\prime(n)=2\zeta(3) \\
&&\sum_{n=1}^\infty\frac{1}{n}\psi^\prime(1+n)=\zeta(3)
\label{s17} \\
&&\sum_{n=1}^\infty\frac{1}{n(n+1)}\psi^\prime(n)=1 \\
&&\sum_{n=1}^\infty\frac{1}{n(n+1)}\psi^\prime(1+n)=-\zeta(3)+\zeta(2)
\end{eqnarray}
\em Proof of Series (\ref{s17}):\em\ We start by using the integral 
representation of the trigamma function:
\begin{eqnarray}
\sum_{n=1}^\infty\frac{1}{n}\psi^\prime(1+n)
&=&-\sum_{n=1}^\infty\frac{1}{n}\int_0^1{\rm d}t\frac{t^n}{1-t}\log t
\nonumber\\[4pt]
&=&\int_0^1\frac{{\rm d}t}{1-t}\log t\log(1-t)\nonumber\\[4pt]
&=&\int_0^1\frac{{\rm d}t}{t}\log t\log(1-t)=\zeta(3)
\nonumber
\end{eqnarray}
The resulting integral is just the integral representation of 
$\mbox{Li}_3(1)$ which equals $\zeta(3)$.
 
Also some series bilinear in $\psi(z)$ can be summed by similar methods:
\begin{eqnarray}
&&\sum_{n=1}^\infty\frac{1}{n(n+1)}[\gamma+\psi(n)]^2=\zeta(2)+1\\
&&\sum_{n=1}^\infty\frac{1}{n(n+1)}[\gamma+\psi(n)][\gamma+\psi(1+n)]
=\zeta(3)+\zeta(2)\\
&&\sum_{n=1}^\infty\frac{1}{n(n+1)}[\gamma+\psi(n)][\gamma+\psi(2+n)]=3 \\
&&\sum_{n=1}^\infty\frac{1}{n(n+1)}[\gamma+\psi(1+n)]^2=3\zeta(3)\\
&&\sum_{n=1}^\infty\frac{1}{n(n+1)}[\gamma+\psi(1+n)][\gamma+\psi(2+n)]
=2\zeta(3)+\zeta(2)\\
&&\sum_{n=1}^\infty\frac{1}{n(n+1)}[\gamma+\psi(2+n)]^2
=\zeta(2)+3
\end{eqnarray}
These series are summed by using the integral representation of 
each of the factors $\gamma+\psi(1+n)$ before summing over $n$. The 
resulting two-dimensional integral yields the desired result. 

Some one-dimensional series containing $\Gamma$-functions also appear.
Consider the following series:
\begin{eqnarray}
\sum_{n=1}^\infty\frac{1}{n^2}\frac{[\Gamma(n)]^2}{\Gamma(2n)}
&=&-\frac{8}{3}\zeta(3) +\frac{4\pi}{3}\Cl{2}\left(\frac{\pi}{3}\right)
\label{gamma-1}\\
\sum_{n=1}^\infty\frac{(-1)^n}{n^2}\frac{[\Gamma(n)]^2}{\Gamma(2n)}
&=&-\frac{4}{5}\zeta(3) \hspace*{10mm}
\end{eqnarray}
\em Proof of (\ref{gamma-1}):\em
\begin{equation}
\begin{split}
&\sum_{n=1}^\infty\frac{1}{n^2}\frac{[\Gamma(n)]^2}{\Gamma(2n)}
=2\sum_{n=1}^\infty\frac{1}{n^2}
\frac{\Gamma\left(\frac{1}{2}\right)\Gamma(n)}
{\Gamma\left(\frac{1}{2}+n\right)}\left(\frac{1}{4}\right)^n\\
&={}_4F_3\left(1,1,1,1;\frac{3}{2},2,2;\frac{1}{4}\right)\\
&=\int_0^1{\rm d}t\,{}_3F_2\left(1,1,1;\frac{3}{2},2;\frac{1}{4}t\right)
\nonumber
\end{split}
\end{equation}
Next, we make use of (7.4.2.353) of \cite{Prudnikov} 
and the substitution 
$\sqrt{t}/2=\sin \frac{u}{2}$ to get
\begin{equation*}
4\int_0^1\frac{{\rm d}t}{t}\arcsin^2\frac{\sqrt{t}}{2}
=2\int_0^{\frac{\pi}{3}}\frac{{\rm d}u}{2\tan\frac{u}{2}}u^2.
\end{equation*}
Integration by parts yields:
\begin{align}
&2\left.u^2\log\left(2\sin \frac{u}{2}\right)\right|_0^\frac{\pi}{3}
-4\int_0^{\frac{\pi}{3}}{\rm d}u\ u\log\left(2\sin \frac{u}{2}\right)
\nonumber \\
&=-4\int_0^{\frac{\pi}{3}}{\rm d}u\ u\log\left(2\sin \frac{u}{2}\right)
=-\frac{8}{3}\zeta(3) +\frac{4\pi}{3}\Cl{2}\left(\frac{\pi}{3}\right)
\nonumber
\end{align}
after using (6.46) and correcting the misprint in (6.52) of \cite{Lewin}.
\section{Two-dimensional series}
In the evaluation of Feynman diagrams by the above outlined methods,
also two- and three-dimensional series are required.
For the purpose of summing certain two-dimensional series,
the following results are very useful:
\begin{equation}
\sum_{n=1}^\infty\frac{1}{n+a}\; \frac{1}{n+b} 
=\frac{1}{b-a}\left[\psi(1+b)-\psi(1+a)\right], \quad a\neq b
\end{equation}
and
\begin{equation}
\sum_{n=1}^\infty\frac{\Gamma(n+k)}{\Gamma(1+n+2k)}
=\frac{\Gamma(k)}{\Gamma(1+2k)} 
\end{equation}

Some selected two-dimensional series that do not involve 
the psi function are:
\begin{eqnarray}
\sum_{n=1}^\infty\sum_{k=1}^\infty\frac{1}{nk(n+k)}&=&2\zeta(3) \\
\sum_{n=0}^\infty\sum_{k=1}^\infty\frac{1}{k(n+k)(1+n+k)}&=&\zeta(2)
\label{s27}  \\
\sum_{n=1}^\infty\sum_{k=1}^\infty\frac{1}{k}
\frac{\Gamma(n)\Gamma(k)}{\Gamma(1+n+k)}&=&\zeta(3) \\
\sum_{n=1}^\infty\sum_{k=1}^\infty\frac{1}{k!}
\frac{\Gamma(2k)\Gamma(n+k)}{\Gamma(1+n+2k)}&=&\half\zeta(2)
\end{eqnarray}
A few typical series involving the psi function are:
\begin{eqnarray}
&&\sum_{n=0}^\infty\sum_{k=1}^\infty\frac{\gamma+\psi(1+k)}{k(n+k)(1+n+k)}
=2\zeta(3) \label{s29}\\
&&\sum_{n=0}^\infty\sum_{k=1}^\infty\frac{\gamma+\psi(1+n)}{k(n+k)(1+n+k)}
=2\zeta(3) \label{s31}\\
&&\sum_{n=0}^\infty\sum_{k=1}^\infty\frac{\gamma+\psi(1+n+k)}{k(n+k)(1+n+k)}
=3\zeta(3) \label{s33}\\
&&\sum_{n=0}^\infty\sum_{k=1}^\infty\frac{\gamma+\psi(1+n+2k)}{k(n+k)(1+n+k)}
=\tfrac{7}{2}\zeta(3) \label{s35}
\end{eqnarray}
\section{Summary}
We have seen that Quantum Field Theory relates to rather interesting
mathematics.
In physical problems of central interest 
(see, e.g., \cite{Groote-Pivovarov,Kalmykov,Melnikov,Petkou,Ritbergen}),
series related to the Euler series emerge.
They have been summed in terms of $\zeta(2)$, $\zeta(3)$ and 
closely related irrational constants.
A computerized treatment of such sums is also available \cite{Vermaseren}.

It is interesting to note that the transcendental basis that is 
required for the considered types of sums is rather restricted.
In particular, there are no $\zeta(4)$, $\zeta(5)$, etc.
The restrictions of such transcendental bases have received a lot
of attention recently, in particular by Broadhurst and Kreimer
\cite{Broadhurst,Kreimer} (see also \cite{Groote}), 
who relate these restrictions to properties of Hopf algebras.

\end{document}